\documentclass[aps,pra,reprint,superscriptaddress,a4paper]{revtex4-2}

\usepackage[utf8]{inputenc}
\usepackage{times}

\usepackage{amsmath, amssymb, amsthm, mathtools, physics, bm, bbold, upgreek, accents}
\usepackage{braket}

\usepackage[pdftex]{graphicx}
\DeclareGraphicsExtensions{.png,.pdf,.eps,.jpg}
\graphicspath{{./figs/}}
\usepackage{subfigure}
\usepackage{array, longtable, multirow, tabu, hhline}
\newcolumntype{M}[1]{>{\centering\arraybackslash}m{#1}} 
\newcolumntype{N}{@{}m{0pt}@{}}

\usepackage{xcolor, color, soul}
\usepackage[normalem]{ulem} 

\usepackage{tikz}
\usetikzlibrary{decorations.pathmorphing, calc, shapes.geometric, positioning, arrows.meta}
\tikzstyle{box} = [draw, fill=blue!10, text centered, rectangle, minimum width=2cm, minimum height=1cm]
\tikzstyle{arrow} = [thick, ->, >=stealth]

\usepackage{tcolorbox}
\usepackage{placeins}
\usepackage{natbib}

\usepackage{hyperref}
\definecolor{myurlcolor}{rgb}{0,0,0.7}
\definecolor{myrefcolor}{rgb}{0.8,0,0}
\hypersetup{
    unicode=true,
    pdfusetitle,
    bookmarks=false,
    breaklinks=false,
    pdfborder={0 0 0},
    backref=false,
    colorlinks=true,
    linkcolor=myrefcolor,
    citecolor=myurlcolor,
    urlcolor=myurlcolor
}

\theoremstyle{plain}

\ifx\proof\undefined
\newenvironment{proof}[1][\proofname]{\par
  \normalfont\topsep6\p@\@plus6\p@\relax
  \trivlist
  \itemindent\parindent
  \item[\hskip\labelsep\scshape #1]\ignorespaces
}{
  \endtrivlist\@endpefalse
}
\providecommand{\proofname}{Proof}
\fi

\newcommand{\eqnref}[1]{Eq.~(\ref{#1})}

\newcommand{\figref}[1]{Fig.~\ref{#1}}

\newcommand{\secref}[1]{Sec.~\ref{#1}}
\newcommand{\appref}[1]{App.~\ref{#1}}

\renewcommand\Tr{\operatorname{Tr}}

\renewcommand{\i}{\mathrm{i}}

\newcommand{\T}{\mathrm{T}}

\newcommand{\E}{\mathcal{E}}

\newcommand{\mrm}[1]{\mathrm{#1}}
\newcommand{\mrmb}[1]{\mathrm{#1}}

\renewcommand{\t}[1]{\text{#1}}

\newcommand{\bpm}{\begin{pmatrix}}
\newcommand{\epm}{\end{pmatrix}}
\newcommand{\beq}{\begin{equation}}
\newcommand{\eeq}{\end{equation}}
\newcommand{\ba}{\begin{align}}
\newcommand{\ea}{\end{align}}
\newcommand{\bi}{\begin{itemize}}
\newcommand{\ei}{\end{itemize}}

\renewcommand{\t}[1]{\tilde{#1}}
\newcommand{\cH}{\mathcal{H}}
\newcommand{\cHnH}{\t{\cH}}
\newcommand{\id}{\mathbb{1}}

\begin{document}

\title{Precision assessment in non-Hermitian systems: A comparative study of three formalisms}
	
	\date{\today}
	
\author{Javid Naikoo}
\email{javid.naikoo@amu.edu.pl}
	\affiliation{Institute of Spintronics and Quantum Information, Faculty of Physics and Astronomy, Adam Mickiewicz University, 61-614 Pozna\'{n}, Poland}
\author{Ravindra W. Chhajlany}
\affiliation{Institute of Spintronics and Quantum Information, Faculty of Physics and Astronomy, Adam Mickiewicz University, 61-614 Pozna\'{n}, Poland}

\author{Jan Ko\l{}ody\'{n}ski}
      \affiliation{Institute of Physics, Polish Academy of Sciences, Aleja Lotnik\'{o}w 32/46, 02-668 Warsaw, Poland}
	\affiliation{Centre of New Technologies, University of Warsaw, Banacha 2c, 02-097 Warszawa, Poland}

\author{Adam Miranowicz}
\affiliation{Institute of Spintronics and Quantum Information, Faculty of Physics and Astronomy, Adam Mickiewicz University, 61-614 Pozna\'{n}, Poland}

\begin{abstract}
	Quantifying measurement precision in quantum systems is vital for advancing quantum technologies such as sensing, communication, and computation. The quantum Fisher information (QFI) sets the ultimate precision bound in Hermitian systems; however, extending this concept to non-Hermitian systems, even those with real spectra, poses conceptual challenges due to their non-unitary dynamics. We compare three probability-conserving approaches for evaluating QFI in such systems: (i) simple normalization, (ii) metric formalism, and (iii) master-equation framework. Although all three ensure probability conservation, they differ in physical interpretation and in how they quantify estimation precision. Our study is particularly motivated by previous studies that have shown that the simple normalization method for non-Hermitian Hamiltonian generated dynamics may lead to misleading or even unphysical conclusions for certain quantum information theoretic tasks.    We emphasize, in this article, that the metric formalism naturally enables the use of standard Hermitian metrology tools in cases where it
    provides a coherent and physically consistent framework for non-Hermitian systems.  
\end{abstract}
	
\maketitle
	
\section{Introduction}
Quantum Fisher information (QFI) is a fundamental concept in quantum metrology that quantifies the amount of information a quantum state encodes about unknown parameters. It serves as a crucial tool for assessing the precision limits of parameter estimation by providing ultimate preecision bound known as  the quantum Cram\'er--Rao bound \cite{Helstrom1967MinimumME,bures1969extension,Helstrom1976}. Mathematically, QFI is derived from the symmetric logarithmic derivative \cite{HelstromSLD1968}, which characterizes how the quantum state evolves with respect to small changes in the parameter of interest \cite{Braunstein1994, Braunstein1996}.  QFI has  been applied to determine limits on the optimal estimation of different quantities, such as phases \cite{Ballester2004,Monras2006,Aspachs2009,Demkowicz2009,Demkowicz2012,Chin2012,Humphreys2013,Nusran2014,Sparaciari2015,Pezze2017}, temperature \cite{Monras2011,Correa2015,Spedalieri2016,Hofer2017,Mehboudi2022}, magnetic fields \cite{Wasilewski2010,Cai2013,Zhang2014,Nair2016,Julia2021}, squeezing parameters \cite{Gaiba2009,Safranek2016}, and so on.

Beyond its role in parameter estimation, QFI has emerged as a versatile tool in the analysis of quantum states and dynamics. It has been extensively used to characterize entanglement \cite{Pezze2009, Toth2012}, quantifying non-Markovianity \cite{QFIflow}, understanding quantum speed limits \cite{Taddei2013},  the quantum Zeno effect \cite{Smerzi2012,Schafer2014} and macroscopic quantum phenomena \cite{FrowisDur2011}. 

Recently, interest has shifted towards non-Hermitian systems, particularly those exhibiting parity-time ($\mathcal{PT}$) symmetry, where unusual physical properties emerge. These systems, unlike Hermitian ones, can have real eigenvalue spectra under specific conditions, despite not conserving probability~\cite{Bender1998}. A notable feature of these systems is the presence of exceptional points, which are singularities in parameter space where two or more eigenvalues and their corresponding eigenvectors coalesce~\cite{Kato1966}. Unlike conventional Hermitian systems, where eigenvalues remain distinct under small perturbations, the emergence of exceptional points in non-Hermitian dynamics leads to a breakdown of the usual eigenvector orthogonality and introduces striking physical consequences~\cite{Ozdemit2019,ElGanainy2018,Parto2020,Heiss2012,Javid2019}.

While traditional quantum metrology has largely focused on Hermitian systems \cite{Degen2017}, the exploration of non-Hermitian systems has opened new pathways for developing more sensitive quantum sensors. Recent studies have demonstrated that $\mathcal{PT}$-symmetric quantum sensors, which operate near exceptional points, can achieve enhanced precision in parameter estimation \cite{Wiersig2014,Wiersig2020}. These systems are particularly interesting because they provide novel mechanisms to surpass classical limits, especially in noisy environments \cite{Chen2017}. However, such enhancement in precision becomes dubious in  regimes where quantum noise becomes significant \cite{Lau2018,Chen2019,Zhang2020,Naikoo2023,Naikoo2025}. 

The QFI in Hermitian systems is calculated with respect to the system state given by a valid density matrix that satisfies the important requirement of probability conservation. In contrast, non-Hermitian systems introduce several complexities. The eigenvalues of a non-Hermitian Hamiltonian (NHH) are generally complex, and the eigenstates may lose orthogonality, necessitating the introduction of a biorthogonal framework where right and left eigenstates are treated on an equal footing \cite{Brody2013, Mostafazadeh2002}.  Therefore, in such systems, QFI cannot be straightforwardly calculated using the traditional Hermitian formalism. The proper formalism of dealing with  non-Hermitian systems with real spectrum involves introducing a metric formalism that adjusts the inner product structure of the Hilbert space, leading to physically meaningful probability-conserving dynamics \cite{Mostafazadeh2010}. This adjustment is necessary, as the non-Hermiticity modifies how quantum probabilities and overlaps are measured in these systems. 

Contemporary research in non-Hermitian quantum metrology have also revealed new pathways for enhanced sensitivity particularly by leveraging \emph{postselection} techniques to engineer effective non-Hermitian dynamics within a larger Hermitian framework \cite{Yaoming2020,Xiao2024,Xinglei2023,Xinglei2024}. While such methods can significantly amplify parameter sensitivity, they also introduce critical considerations regarding the trade-off between QFI and the probability of successful postselection \cite{Tanaka2013,Combes2014}. Since postselection inherently discards a fraction of experimental outcomes, the effective metrological advantage must be assessed in terms of the total Fisher information per experimental run, rather than the raw QFI of the selected subensemble. Consequently, the widespread practice of wavefunction normalization after non-unitary evolution may lead to an overestimation of QFI by keeping only the most favorable trajectories, thereby ignoring the total statistical cost of the full process. These observations underscore the importance of careful accounting of probability conservation in assessing the metrological potential of non-Hermitian systems, thereby ensuring that reported enhancements translate into tangible experimental advantages.

 The motivation for this work stems from the observation that, whilst quantum metrology is well developed within the framework of conventional quantum mechanics based on  probablility conservation that is guarenteed by fundamental equations of motion and measurement theory (projective and more generally positive-operator valued measures), non-Hermitian metrology can be addressed in non-unique and ways not necessarily consistent with all the postulates of quantum mechanics and physical principles. 

Recall that Hermitian Hamiltonians $ H = H^\dagger$ with parameters  $ \Omega^\mrm{R}_{\mrm{H}}$ that correspond to real spectra and thus generate unitary dynamics which describe closed (or isolated) systems.  The NHHs where $ H \neq H^\dagger $, appear most naturally in open quantum systems and effective or approximate descriptions of dissipation. In the regime where at least one eigenvalue becomes \emph{complex}, the system belongs to the region $ \Omega^\mrm{C}_{\mrm{NH}} $, defined by  
\begin{equation}
	\exists~ \lambda_i \notin \mathbb{R}, \quad \text{s.t.} \quad H \psi_i = \lambda_i \psi_i.
\end{equation}
In this case, the system evolution is inherently non-unitary in the standard Hilbert space, and metrological studies typically rely on quantum master equation approaches for a consistent treatment.

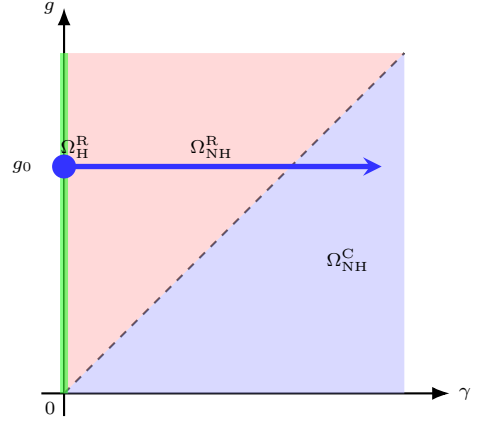
\begin{figure}[t]
	\centering
	\begin{tikzpicture}[scale=1.5, font=\scriptsize]
		\begin{scope}[xshift=6cm]
			\draw[thick,-Latex] (-0.2,0) -- (3.4,0) node[right] {$\gamma$};
			\draw[thick,-Latex] (0,-0.2) -- (0,3.4) node[left] {$g$};
			
			\node[below left] at (0,0) {$0$};
			
			\draw[thick, dashed] (0,0) -- (3,3);
			
			\begin{scope}
				\clip (0,0) -- (3,3) -- (0,3) -- cycle;
				\fill[red!30, opacity=0.5] (0,0) rectangle (3,3);
			\end{scope}
			
			\begin{scope}
				\clip (0,0) -- (3,3) -- (3,0) -- cycle;
				\fill[blue!30, opacity=0.5] (0,0) rectangle (3,3);
			\end{scope}
			
			\draw[line width=3pt, green!80, opacity=0.6] (0,0) -- (0,3.0);
			
			\draw[thick,->,>=stealth, color=blue!80, line width=2pt] (0,2) -- (2.8,2);
			
			\fill[blue!80] (0,2) circle (3pt);
			
			\node[left] at (-0.2,2) {$g_0$};
			
			\node[above] at (0.1,2) {$\Omega_\mrm{H}^\mrm{R}$};
			\node[above] at (1.3,2) {$\Omega_\mrm{NH}^\mrm{R}$};
			\node[above] at (2.5,1) {$\Omega_\mrm{NH}^\mrm{C}$};
			
		\end{scope}
	\end{tikzpicture}
	\caption{
		\textbf{Different possible regions in the parameter space} of a specific Hamiltonian $H = \begin{bmatrix} -\i \gamma & g \\ g & \i \gamma \end{bmatrix}$, illustrating the transition along the slice $g=g_0$. The system moves from $\Omega_\mrm{H}^\mrm{R}$ at $\gamma = 0$ to $\Omega_\mrm{NH}^\mrm{R}$ for $0 < \gamma < g$, and then to $\Omega_\mrm{NH}^\mrm{C}$ for $0 < g < \gamma $. Here, $\Omega_{\rm H}^{\rm R}$ and $\Omega_{\rm NH}^{\rm C}$ denote regions in the parameter space of $H$ corresponding to Hermitian Hamiltonians with real eigenvalues and NHHs with complex eigenvalues, respectively, while $\Omega_{\rm NH}^{\rm R}$ refers to the intermediate region of NHHs that still possess real eigenvalues.}
	\label{fig:parameter_spaces}
\end{figure}

An intermediate regime, $ \Omega^\mrm{R}_{\mrm{NH}} $, consists of NHHs that nevertheless have entirely \emph{real} eigenvalues:  
\begin{equation}
	H \neq H^\dagger, \quad \lambda_i \in \mathbb{R} \quad \forall i.
\end{equation}
Here, a suitable choice of an inner-product structure, often referred to as a ``metric" formalism, allows the system to retain unitary-like evolution under a redefined notion of unitarity. Specifically, there exists a positive-definite operator  $\eta$ (the so called metric) such that the Hamiltonian is pseudo-Hermitian
\begin{equation}
	H\in \Omega^\mrm{R}_{\mrm{NH}} 
	\quad \Leftrightarrow \quad
	\exists\,\eta > 0\;\;\text{s.t.}\;\; H^{\dagger} = \eta H \eta^{-1}.
	\label{eq:nonHerm_Ham}
\end{equation}
Indeed this central result can be paraphrased upon closer inspection as  "pseudo-hermitian Hamiltonians  are Hermitian in disguise” as will be presented in Sec.\ref{sec:approaches}.
Note that as $\Omega^\mrm{R}_{\mrm{H}}\subset \Omega^\mrm{R}_{\mrm{NH}}$, the Hermitian case corresponds to the special situation when  $\eta=\id$. The qualitative structure of these regions in a parameter space is illustrated in~\figref{fig:parameter_spaces}, which shows how the system transitions across $\Omega^\mrm{R}_{\mrm{H}} $, $\Omega^\mrm{R}_{\mrm{NH}} $, and $\Omega^\mrm{C}_{\mrm{NH}}$ for a representative two-level Hamiltonian. More broadly, using   more general time-dependent metrics may also allow mapping of non-Hermitian Hamiltonians with complex eigevalues to Hermitian Hamiltonians. In both these cases, where mapping to Hermitian Hamiltonians is possible, the system in an appropriate reference frame realizes exact closed system dynamics.

In this work, we adopt the well established metric formalism to demonstrate the locally Hermitian approach to quantum metrology in $ \Omega^\mrm{R}_{\mrm{NH}} $. Rather than claiming enhanced precision or a metrological advantage, we show how standard quantum metrology concepts should be correctly adapted within this formalism. Our results provide a well-founded framework for analyzing parameter estimation in this regime, ensuring consistency with the broader principles of quantum mechanics. We also compare this approach with the no-jump evolution, where the state is normalized at each time step, and also with the results obtained using the full master equation. These comparisons elucidate the advantages and limitations of different methodologies in non-Hermitian quantum metrology and offer insights into optimizing estimation strategies.

This paper is organized as follows. Section~\ref{sec:approaches} examines various approaches to quantum dynamics in the presence of a NHH. Section~\ref{sec:QFI_NHH} discusses QFI in non-Hermitian systems in the context of metric formalism. In particular, we explore how both Hamiltonian and state parameters can be inferred  from the NHH and the Schrödinger equation. Section~\ref{sec:PhysSystems} investigates two distinct physical scenarios in which a qubit undergoes open system dynamics and the underlying metric is time-independent in one case and time-dependent in the other, illustrating different methods for estimating state parameters by computing QFI and comparing their effectiveness. Finally, ~\secref{sec:Conclusion} summarizes our findings.

\section{Approaches to Non-Hermitian Dynamics} \label{sec:approaches}
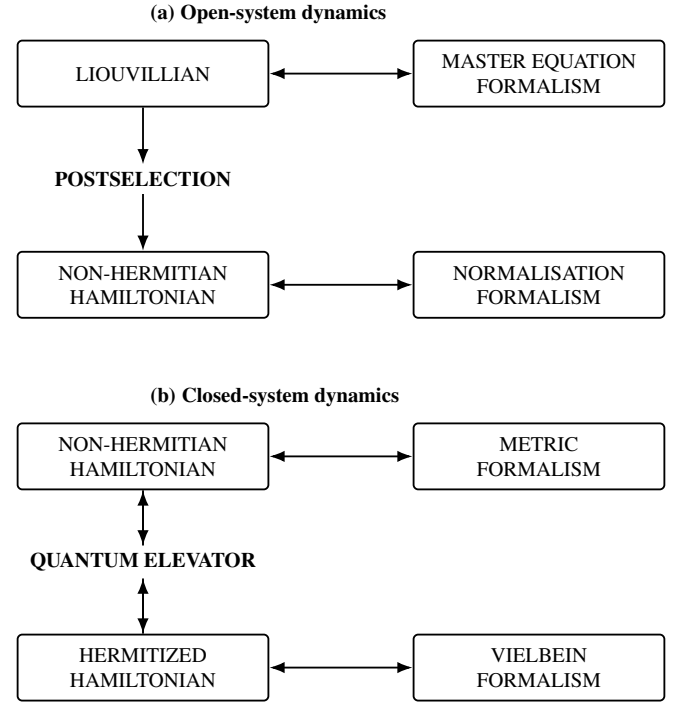
\begin{figure}[t]
\centering
\resizebox{\columnwidth}{!}{%
\begin{tikzpicture}[>=Latex]

\tikzset{
  box/.style={draw, rounded corners=2pt, thick,
              minimum width=3.8cm, minimum height=1.0cm,
              align=center},
  labeltext/.style={font=\small\bfseries},
  arrow/.style={->, thick},
  doublearrow/.style={<->, thick}
}

\node[labeltext, anchor=west] (titlea) at (0,0) {(a)\;Open-system dynamics};

\begin{scope}[shift={(0,-0.9)}]
  \node[box] (liou) at (0,0) {LIOUVILLIAN};
  \node[box] (mef)  at (6,0) {MASTER EQUATION\\FORMALISM};
  \draw[doublearrow] (liou) -- (mef);

  \node[labeltext] (post) at ($(liou)+(0,-1.6)$) {POSTSELECTION};
  \draw[arrow] (liou) -- (post);

  \node[box] (nh)   at ($(post)+(0,-1.6)$) {NON-HERMITIAN\\HAMILTONIAN};
  \node[box] (norm) at ($(nh)+(6,0)$) {NORMALISATION\\FORMALISM};
  \draw[doublearrow] (nh) -- (norm);

  \draw[arrow] (post) -- (nh);
\end{scope}

\node[labeltext, anchor=west] (titleb) at (0,-5.8) {(b)\;Closed-system dynamics};

\begin{scope}[shift={(0,-6.7)}]
  \node[box] (nh2)   at (0,0) {NON-HERMITIAN\\HAMILTONIAN};
  \node[box] (norm2) at (6,0) {METRIC\\FORMALISM};
  \draw[doublearrow] (nh2) -- (norm2);

  \node[labeltext] (elev) at ($(nh2)+(0,-1.6)$) {QUANTUM ELEVATOR};
  \draw[arrow] (nh2) -- (elev);
  \draw[arrow] (elev) -- (nh2);

  \node[box] (herm)   at ($(elev)+(0,-1.6)$) {HERMITIZED\\HAMILTONIAN};
  \node[box] (metric) at ($(herm)+(6,0)$) {VIELBEIN\\FORMALISM};
  \draw[doublearrow] (herm) -- (metric);

  \draw[arrow] (elev) -- (herm);
  \draw[arrow] (herm) -- (elev);
\end{scope}

\end{tikzpicture}
}
\caption{Types of dynamics  and their mutual relationships (left),  and correponding formalisms (right). (a) Note that the master-equation formalism corresponds to the
standard quantum-trajectory approach without post-selection, i.e.,
allowing an arbitrary number of quantum jumps, whereas the
normalization formalism corresponds to the quantum-trajectory
approach post-selected on no quantum jumps. Post-selection in open systems (Liouvillian) dynamics leads to non-Hermitian effective Hamiltonians. (b) Given a non-Hermitian Hamitlonian,  in an appropriately transformed  reference frame, dubbed ``Einstein's quantum elevator" following \cite{Ju2022}, a non-Hermitian Hamiltonian is rendered Hermitian. An equivalent description of the physical system is obtained via the metric formalism.}
\label{fig:formalisms}
\end{figure}

This section  introduces key formalisms--the metric formalism, the normalization method, and the master equation approach--highlighting their roles in governing differing quantum state evolutions corresponding to the same non-Hermitian Hamiltonian. These frameworks will later be used to analyze the precision of estimating unknown state parameters and to examine the resulting differences in  behavior of the precision across the formalisms. A visual depiction summarizing the relationships of these descriptions is provided in Fig.~\ref{fig:formalisms}.

\subsection{Metric framework}  

The Schr\"odinger equation with a non-Hermitian Hamiltonian leads to non-unitary evolution, \textit{viz.}  the non-conservation of the standard (i.e. familiar Euclidean) inner product $\langle\cdot|\cdot\rangle$ during evolution. 
A key idea of consistent non-Hermitian quantum mechanics is that certain non-Hermitian operators (e.g. those with real eigenvalues dubbed pseudo-Hermitian \cite{Mostafazadeh2002})  preserve modified inner products $\langle\cdot|\cdot\rangle_\eta:=\langle \cdot|\eta(t) |\cdot\rangle$, where $\eta(t)$ is a (possibly time-dependent) so-called metric operator, which is Hermitian and positive definite. Thus such non-Hermitian operators generate unitary evolution in modified Hilbert space with inner product $\langle\cdot|\cdot\rangle_\eta$. Moreover, the metric operator and Hermitian action of the considered Hamiltonian can be equivalently effected implicitly as a change of reference frame of the  system via a general, non-unitary yet invertible transformation. The latter may be considered a generalization of the concept of Einstein's elevator in relativity to quantum mechanics \cite{Ju2022} - a system in a space with a non trivial metric in an inertial frame of reference can be viewed equivalently as a system with a trivial metric in a non-inertial frame. We illustrate these points below.

Consider a system $\mathcal{S}_\mrm{NH}$ described by a NHH $\t{H} : \cHnH \to \cHnH$. The non-Hermiticity implies $\t{H}^\dagger \ne \t{H}$ where the adjoint $^\dagger$ is defined by the relation $\langle \t{H}^\dagger \t{\psi} | \t{\psi} \rangle = \langle \t{\psi} | \t{H} \t{\psi} \rangle$ and $\langle \t{\psi}  | \t{\phi}  \rangle$ is the \emph{standard} inner product. 

The Schr\"odinger equation describing this system is
\begin{align}
    \i \frac{d}{dt} \ket{\t{\psi}(t)} &= \t{H} \ket{\t{\psi}(t)} \quad {\rm non\!\!-\!\! Hermitian~system},\mathcal{S}_\mrm{NH} 
\end{align}
Let $\cH_{\eta}$ be the space $\cHnH$  equipped with the metric defined by the positive-definite Hermitian operator $\eta$. In order to preserve the modified inner product under the above dynamics ~\cite{Mostafazadeh2020TIme,ChiaYi2022}:
	\begin{equation}
\i\frac{d}{dt}\langle \t{\psi} |\t{\phi}\rangle_\eta =\frac{d}{dt}\langle \t{\psi}| \eta |\t{\phi}\rangle = 0	\Rightarrow 	i \frac{d}{d t} \eta(t) = \t{H}^{\dagger} \eta(t) - \eta(t) \t{H}.
		\label{eq:metricEOM}
	\end{equation}
Although we only consider time independent Hamiltonians in this Article, the above condition is valid for time dependent Hamiltonians. We note here that in the particular case of pseudo-Hermitian Hamiltonians, the metric can be chosen constant in time since the above equation is simply satisfied by the pseudo-Hermiticity condition $\t{H}^{\dagger} = \eta \t{H} \eta^{-1}$.  More generally, however,  metric solutions to Eq.~\eqref{eq:metricEOM} may also exist  in cases where the NHH has complex eigenvalues (see Sec.~\ref{subsec:tdm}).

The metric formalism may be reformulated as a transformation to a different basis mapping the non-Hermitian system to a Hermitian one with a standard inner product (with trivial metric). Indeed, note firstly that a positive definite Hermitian operator can be decomposed as $\eta = \E^\dagger \E$. The metric modified scalar product then is 
	\begin{align}
		\langle \t{\psi} |\t{\phi} \rangle_{\eta}  = 	& \langle \t{\psi}| \eta |\t{\phi} \rangle = \langle \t{\psi}| \E^\dagger \E |\t{\phi} \rangle
				:= \langle \psi | \phi \rangle,
	\end{align}
where the role of $\E$ effects the state transformation $  |\psi\rangle = \E |\t\psi \rangle $. These  states in the transformed basis evolve according to the Schr\"odinger equation 
\begin{align}
\i \frac{d}{dt} \ket{\psi(t)} &= h \ket{\psi(t)} \quad {\rm ~Hermitized~system},\;\mathcal{S}_\mrm{H}
\end{align}
with Hermitian Hamiltonian 
\begin{equation}
	h = \E \t{H} \E^{-1} + \i  (\partial_t \E) \E^{-1}.	
	\label{eq:Herm_Ham}
\end{equation}
The evolution of the system in this frame of reference is then manifestly unitary.

The transformation $\E$ which can be called  a generalized vielbein ~\cite{ChiaYi2022,Ju25} (see \appref{app:vielbein} for a short recap of vielbein's as used in general relativity) and we call the above Hermitization the vielbein formalism. Intuitively, the vielbein provides a local mapping that preserves the geometry encoded by $\eta$, allowing one to work in a familiar Hilbert space while fully accounting for the effects of the modified inner product. The practical utility of the Hermitized frame is that  it allows the use of standard quantum theory, including notions of probabilities and measurement theory, consistently.
   	\begin{align}
		\frac{\langle \t{\psi}  | \t{A} | \t{\psi} \rangle_{\eta} }{\langle \t{\psi} | \t{\psi} \rangle_{\eta} } 
		= \frac{\langle \t{\psi}  | \eta \t{A}  | \t{\psi} \rangle}{\langle \t{\psi} | \eta| \t{\psi} \rangle} 
	=  	\frac{\langle \t{\psi} | \E^\dagger  \E \t{A} | \t{\psi} \rangle }{\langle \t{\psi} | \E^\dagger \E | \t{\psi} \rangle } 
		\\ =	\frac{\langle \t{\psi} | \E^\dagger  \E \t{A} \E^{-1} \E | \t{\psi} \rangle }{\langle \t{\psi} | \E^\dagger \E | \t{\psi} \rangle }
		= \frac{\langle \psi|  A  | \psi \rangle }{\langle \psi | \psi \rangle },
		\label{eq:avg}
	\end{align}
where the observable $A$ is Hermitian and $\t{A} = \E^{-1} A  \E $. Note that, for convenience of comparison with the expressions for analogous expressions in the normalization formalism, we explicity divide by the norm of the states which are however constant by construction, \textit{i.e.}
$\eta$ is chosen such that $\langle \tilde{\psi} | \eta |\tilde{\psi} \rangle=\langle \psi | \psi \rangle=  1 $ is normalized, while  $\langle \tilde{\psi}|\tilde{\psi} \rangle \neq 1$, in general.

\subsection{Normalization approach}\label{sec:NormApproach}
When a NHH  $\tilde{H}$ is employed within the Schrödinger equation,  the evolution induced by $\tilde{H}$ is not trace-preserving, rendering the state physically inconsistent. A standard procedure to address this issue is to \emph{renormalize} the state at each instant by its trace. For an arbitrary initial state $\rho_0$, this prescription reads
\begin{equation}
	\rho_\mathrm{norm}(t) = \frac{e^{-\i \tilde{H} t} \rho_0 \, e^{\i \tilde{H}^\dagger t}}{\mathrm{Tr}\big[ e^{-\i \tilde{H} t} \rho_0 \, e^{\i \tilde{H}^\dagger t} \big]},
\end{equation}
thereby restoring unit trace and ensuring the physical validity of the state. The normalized physical quantities are then computed as:
\begin{equation}
	\langle A(t) \rangle_\mrm{norm} = \frac{\langle \tilde{\psi}(t) | A | \tilde{\psi}(t) \rangle}{\langle \tilde{\psi}(t) | \tilde{\psi}(t) \rangle} = \frac{\Tr[\tilde{\rho}(t) A]}{\Tr[\tilde{\rho}(t)]} = \Tr[\rho_\mrm{norm}(t) A]  .\label{eq:Anorm}
\end{equation}
Note the fundamental difference in the expression for the  averatge value of the observable $A$ in the above normalization form and previously mentioned metric formulation Eq.~\eqref{eq:avg}.

Interestingly, $\rho_\mrm{norm}(t)$ satisfies the following non-linear equation of motion~\cite{Brody2012}
\begin{align}
	&\frac{d}{dt} \rho_\mrm{norm}(t) = -\i\, [\t{H}_\mrm{H}, \rho_\mrm{norm}(t)]  \\
	&\qquad- \left( \{\t{H}_\mrm{S}, \rho_\mrm{norm}(t)\}  - 2 \Tr[\rho_\mrm{norm}(t) \t{H}_\mrm{S}] \rho_\mrm{norm}(t) \right),\nonumber
\end{align}
where $\t{H}_\mrm{H}$ and $\t{H}_\mrm{S}$ are the Hermitian and skew-Hermitian parts of $\t{H} = \t{H}_\mrm{H} - \i \t{H}_\mrm{S}$, and $\{\cdot,\cdot\}$ denotes the anticommutator. 
 
The normalization procedure is widely used in systems governed by effective NHH. However, despite its mathematical convenience, this approach warrants careful examination, as it may obscure the true dynamics and physical interpretation of the system’s evolution. This is a naive approach that, as discussed in the next section, neglects quantum jumps in systems governed by a non-Hermitian Hamiltonian while still assuming a trivial metric identical to that used in Hermitian systems. As explained in the Introduction, neglecting the proper	non-trivial metric---which should be specifically constructed for 	a given NHH  and used consistently ---can lead to apparent violations of fundamental no-go
	theorems, as reported in various publications (see, e.g., 	\cite{Lee2014, Chen2014, Pati2014, Dogra2021}). However, as emphasized
	previously, a proper inclusion of the correct metric prevents any 	such violations \cite{Znojil2016, Brody2016, Ju2019}.
	
	To illustrate the nature of this issue, we quote a strong 	criticism of the naive-metric approach expressed in \cite{Znojil2016}, 	where it is described as being based on ``defining the 	Hilbert-space structure in a false and manifestly unphysical, 	ill-chosen, and purely auxiliary friendlier space.''
	
	Nevertheless, it is quite remarkable that this---strictly speaking, unphysical  without postselection---approach often yields excellent agreement 	with more rigorous calculations based on the master equation or 	Heisenberg-Langevin equations of motion, at least for systems with 	a small number of excitations, where quantum jumps do not play a 	significant role in the system's dynamics. This situation occurs, 	for example, in photon blockade (PB), photon-induced tunneling, 	and related effects described by second-order or higher-order 	photon-number correlation functions.
	
	By analyzing the results presented in the following references, one can observe a very good correspondence between the analytical 	results obtained within this naive NHH approach and the numerical 	results from the master-equation framework. These examples include 	conventional \cite{Wang2015} and unconventional \cite{Bamba2011, Tang2025} PB, nonreciprocal conventional PB \cite{Huang2018} and its 	unconventional counterpart \cite{Li2019}, chiral PB \cite{Zuo2024}, as well 	as related effects involving the suppression of mechanical 	excitations, i.e.: phonon blockade \cite{Wang2016} and hybrid 	photon-phonon blockade \cite{Abo2022} among others.
	
	For this reason, we also consider this simple NHH approach in the present work---at least for low-excitation regimes, where it might 	provide reliable and physically meaningful results despite its conceptual limitations.

\subsection{Conditional non-Hermitian dynamics in master equations}  \label{sec:MasterEqn}
Dissipation is a key source of non-Hermiticity in quantum systems. Under the Born-Markov approximation, open quantum systems are governed by the master equation in the Lindblad form:
\begin{align}
	\dot{\rho}_\mrmb{me}(t) &=   \mathcal{L} \rho_\mrmb{ me}(t) = -\i\, [H, \rho_\mrmb{me}(t)] \nonumber \\&+ \sum_{\ell} \left( \Gamma_{\ell} \rho_\mrmb{me}(t)  \Gamma_{\ell}^\dagger - \frac{1}{2}  \{ \rho_\mrmb{me}(t), \Gamma_{\ell}^\dagger \Gamma_{\ell} \}\right),  \label{eq:ME}
\end{align}
where \(\rho_\mrmb{me}\) is the system's density matrix,  $\mathcal{L}$ is the Liouvillian superoperator,  \(H\) is the Hamiltonian and \(\Gamma_j\) are Lindblad operators describing the system-environment interaction. For a comparative analysis of
Liouvillian spectra with and without quantum jumps in the context
of exceptional points, see~\cite{Minganti2019, Minganti2020}. The expectation value of an observable $A$ is given by $\langle A \rangle_\mrmb{me} = \text{Tr}(\rho_\mrmb{me} A)$. Directly solving the master equation is computationally intensive. Instead, the quantum trajectory method tracks individual quantum states evolving under an effective NHH
\begin{equation}\label{eq:Heff}
	H_\mrmb{eff} = H - \frac{\i}{2} \sum_{\ell}  \Gamma^\dagger_{\ell} \Gamma_{\ell},
\end{equation}
 punctuated by random quantum jumps. Averaging over the ensemble of pure states reproduces the density matrix \(\rho_\mrmb{me}\), and observables can be calculated accordingly.

The conditional evolution under the assumption that ``no jump" has occurred is described by
\begin{equation}
	\dot{\rho}_\mrmb{nj}(t) = - \i\, (H_\mrmb{eff} \rho_\mrmb{nj}(t) - \rho_\mrmb{nj}(t) H_\mrmb{eff}^\dagger), \label{eq:rhonj}
\end{equation}
neglecting the quantum jump term. The corresponding expectation value  is given by
\begin{equation}
	\langle A(t) \rangle_\mrmb{nj} = \frac{\Tr[\rho_\mrmb{nj}(t) A]}{\Tr[\rho_\mrmb{nj}(t)]}. \label{eq:Anj}
\end{equation}
It should be emphasized that Eq. \eqref{eq:Anj} correspond respectively to Eq. (\ref{eq:Anorm}) discussed in  normalization approach above, wherein one considers only those experimental realizations in which no quantum jumps occur. This \emph{postselection} is inherently optimistic because it neglects the effects of quantum jumps, which typically induce decoherence and diminish the precision of parameter estimation. Despite its theoretical appeal, post-selected quantum metrology inherently discards a large fraction of data, rendering any gain in QFI ineffective due to the low success probability \cite{Combes2014}.

Alternatively, another fully quantum approach based on the Heisenberg-Langevin (HL) equations of motion could be considered. This method has garnered increasing interest recently, especially in analyzing quantum exceptional points of Gaussian fields [see Refs.~\cite{Perina2022Quantum, Perina2023Unavoidability, Naikoo2023, Wakefield2024, 	Perina2024Multiple, Thapliyal2024Multiple}]. Although the HL approach yields predictions consistent with the master equation approach, it is not described in detail here. For a detailed comparison of the normalization and HL approaches for Gaussian processes, see Ref.~\cite{Perina2023Unavoidability}.
 \begin{figure} 
 	\centering
 	\resizebox{\columnwidth}{!}{ 
 	\begin{tikzpicture}[
 		node distance=0.8cm and 0.8cm,  
 		box/.style = {draw, minimum width=2.8cm, minimum height=0.8cm, align=center, font=\large, line width=1.5pt},  
 		arrow/.style = {draw, -{Latex[length=1.5mm]}, line width=1.5pt}  
 		]
 		
 		\node (initial) [box] {Initial State \\ $\rho_\mathrm{in}(\theta)$};
 		
 		\node (rho1) [box, below left=of initial, xshift=-1.2cm] {
 			$\tilde{\rho}(t) = e^{-i \tilde{H} t} \rho_\mathrm{in}(\theta) e^{i \tilde{H}^\dagger t}$ \\
 			$\tilde{\rho}_\mathrm{norm}(t) = \frac{\tilde{\rho}(t)}{\Tr[\tilde{\rho}(t)]}$
 		};
 		\node (Fnorm) [box, below=of rho1, yshift=-0.3cm] {$\mathcal{F}_\theta^\mathrm{norm}$};
 		
 		\node (rho2) [box, below=of initial] {
 			$\rho_\mathrm{metric}(t) = e^{-iht} \rho_\mathrm{in}(\theta) e^{iht}$ \\ 
 			$ h = \mathcal{E} \tilde{H} \mathcal{E}^{-1}  + i (\partial_{t}\mathcal{E}) \mathcal{E}^{-1}$ 
 		};
 		\node (Fmetric) [box, below=of rho2, yshift=-0.3cm] {$\mathcal{F}_\theta^\mathrm{metric}$};
 		
 		\node (rho3) [box, below right=of initial, xshift=1.2cm] {
 			$\rho_\mathrm{me}(t) = \Lambda_{t}[\rho_\mathrm{in}(\theta)]$ \\
 			$\tilde{H} = H_\mathrm{eff}$
 		};
 		\node (Fme) [box, below=of rho3, yshift=-0.3cm] {$\mathcal{F}_\theta^\mathrm{me}$};
 		
 		\draw [arrow] (initial) -- (rho1);
 		\draw [arrow] (rho1) -- (Fnorm);
 		
 		\draw [arrow] (initial) -- (rho2);
 		\draw [arrow] (rho2) -- (Fmetric);
 		
 		\draw [arrow] (initial) -- (rho3);
 		\draw [arrow] (rho3) -- (Fme);
 	\end{tikzpicture}
 	} 
 	\caption{ \textbf{Different ways of evaluating the QFI} given an initial state $\rho_\mrm{in}(\theta)$ and a NHH $\tilde{H}$ generating the evolution. The approaches include: (1) normalizing the non-Hermitian state $\tilde{\rho}(t)$ by its trace at each time step, (2) transforming the state into a metric space with a modified inner product, (3) solving the full master equation to incorporate both coherent evolution and dissipation in open systems. The reason we compare these approaches is that they all describe probability-conserving dynamics.}
 	\label{fig:QFIflowchart}
 \end{figure}
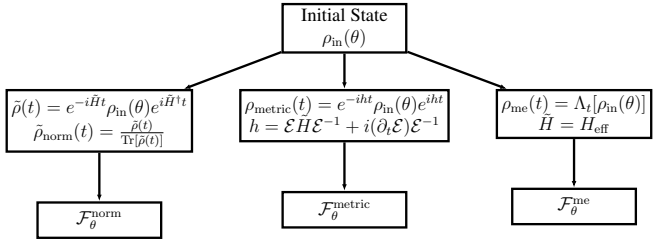

\section{ QFI for non-Hermitian Hamiltonians}\label{sec:QFI_NHH}
  In parameter estimation, the goal is to determine an estimator $\hat{\theta}$  that maps measurement outcomes $\chi$ to a parameter space. In classical estimation theory, an estimator is considered optimal if it achieves the Cramér-Rao bound $\delta^2 \theta  \geq 1/\nu F_{\theta}$, where $\delta^2 \theta$ denotes the mean square error, $\nu$ is the number of measurements, and $F_{\theta}$ is the Fisher information  given by \cite{KayBook}:
\begin{align}
	F_{\theta} =& \int dx \, p(x|\theta) \left(\frac{\partial \ln p(x|\theta)}{\partial \theta}\right)^2, \nonumber \\&= \int \frac{1}{p(x|\theta)} \left(\frac{\partial p(x|\theta)}{\partial \theta}\right)^2 dx.
   \end{align}
For unbiased estimators, the mean square error is equivalent to the variance $\text{Var}(\theta) = E_\theta[\hat{\theta}^2] - E_\theta[\hat{\theta}]^2$.

In quantum parameter estimation, one deals with a family of quantum states $\rho_\theta$ depending on unknown parameter $\theta$. The parameter $\theta$ is inferred through generalized measurements on $\rho_\theta$ and the probability distribution of outcomes is expressed as  $p(x|\theta) = \text{Tr}[\Pi_x \rho_\theta]$, where $\{\Pi_x\}$ are the elements of a POVM (Positive-Operator Valued Measure) with $\int dx \Pi_x = \mathbb{1}$.  The classical Fisher information in this  context is \cite{Braunstein1994}:
\begin{align}
	F_{\theta} =& \int dx \,  \frac{\Tr\left[  \Pi_{x} \partial_{\theta} \rho_{\theta} \right]^2 }{\Tr\left[\rho_{\theta} \Pi_x \right]}, \nonumber \\&=  \int dx \, \frac{\left[\text{Re}(\text{Tr}[\rho_\theta \Pi_x L_\theta])\right]^2}{\text{Tr}[\rho_\theta \Pi_x]},  \label{eq:FI_Def}
\end{align}
where $L_{\theta}$ is the symmetric logarithmic derivative and is defined implicitly as  $\partial_{\theta} \rho_{\theta} = (L_{\theta} \rho_{\theta} +  \rho_{\theta} L_{\theta})/2$.  Maximizing this bound over all possible  quantum measurements (POVMs) determines the ultimate attainable limit of precision in terms of the QFI $\mathcal{F}_{\theta}$ such that 
\begin{equation}
	F_{\theta}  \le \mathcal{F}_{\theta} =  \Tr\left[ \rho_{\theta}  L_{\theta}^2\right].  \label{eq:QFI_EM}
\end{equation}
In the more general setting of multiparameter estimation, \eqnref{eq:QFI_EM} generalises to a matrix inequality that, however, is no longer guaranteed to be saturable despite maintaining its validity~\cite{Rafal2020}. For a given set of parameters $\bm{\theta} = \{\theta_\ell\}_{\ell=0}^n$, each entry of the QFI matrix of size $n\times n$ is then indexed by parameters from the set, and for any $\theta_i,\theta_j\in\bm{\theta}$ it reads
\begin{equation}
	\mathcal{F}_{\theta_i \theta_j} = \frac{1}{2} \text{Tr} (\rho_{\theta} \{L_{\theta_i}, L_{\theta_j}\}). \label{eq:qfim_def}
\end{equation}
where $\{\cdot,\cdot\}$ denotes again the anticommutator.  

For instance, in the special case of a \emph{mixed qubit state} $\rho_{\bm{\theta}}$, each entry of the QFI matrix is given by the following formula~\cite{Shunlong2004,Zhong2013}:
\begin{align}\label{eq:QFImixed}
	\mathcal{F}_{\theta_i, \theta_j} &= \Tr[(\partial_{\theta_i}\rho_{\theta}) (\partial_{\theta_j}\rho_{\theta})] \nonumber \\&+ \frac{1}{\det(\rho_{\theta})} \Tr[\rho (\partial_{\theta_i} \rho_{\theta}) \rho_{\theta} (\partial_{\theta_j} \rho_{\theta})].
\end{align} 

While the QFI is well-defined for Hermitian systems, extending its formulation to non-Hermitian dynamics introduces challenges. The NHHs induce non-unitary evolution, yield complex eigenvalues, and break probability conservation, thereby complicating the direct application of standard QFI formulas. As discussed in  \secref{sec:approaches} and illustrated in \figref{fig:QFIflowchart}, several possibilities exist to tackle these issues, including wavefunction normalization, the metric formalism, and the  master equation approach. Each method has distinct implications for the computation of QFI, and in the following we explore how the QFI behaves depending on the approach undertaken.

In partricular, in the metric formalism, where the inner product is explicitly modfified, an immediate issue is that both probabilities and measurement operators need to be redefined to include the metric operator $\eta(t)$ (similarly to expressions on the l.h.s. of Eq,~\eqref{eq:avg}) seemingly complicating a direct analysis of optimal quantum measurements and thus the QFI. Fortunately, as summarized in Sec.~\ref{sec:approaches}, one may instead equivalently resort to the vielbein formalism, i.e. describe the dynamics in a corresponding Hermitized frame. In the latter frame, the formulae Eqns.~(\ref{eq:FI_Def},\ref{eq:QFI_EM},\ref{eq:qfim_def}), trivially hold and therefore quantify measuremnt precision for the non-Hermitian problem.

However, as each of the approaches summarised in \figref{fig:QFIflowchart} leads to different dynamics given the non-Hermitian $\t{H}$ considered, it would be inappropriate to compare them by considering metrology scenarios in which the parameters $\bm{\theta}$ are encoded by the NHH itself, i.e.~$\t{H}(\bm{\theta})$. Each approach would then yield different $\bm{\theta}$-encoding mechanism, making their corresponding QFIs incomparable. That is why, when dealing with particular physical system in \secref{sec:PhysSystems}  we assume the parameters to be encoded onto the initial state $\rho_\mrm{in}(\bm{\theta})$, as emphasised in \figref{fig:QFIflowchart}. Each approach determines then the $\bm{\theta}$-independent evolution over time $t$ onto the final state $\rho(t)$, whose QFI quantifies then the sensitivity to changes of $\bm{\theta}$ in the initial state common to all three methods.

\begin{figure}
	\centering
	\resizebox{\columnwidth}{!}{ 
		\begin{tikzpicture}[
			node distance=1.5cm and 2.0cm, 
			auto, 
			every node/.style={align=center}, 
			box/.style={draw, rectangle, minimum height=0.8cm, minimum width=1.6cm, rounded corners, font=\footnotesize}, 
			every edge/.style={draw, -{Stealth}, thick}
			]
			
			\node[box] (S) {$\mathcal{S}_\mrm{NH}$\\Non-Hermitian System  };
			\node[box, right=of S] (S') {$\mathcal{S}_\mrm{H}$\\Hermitian System};
			\node[circle, draw, above=1.2cm of $(S)!0.5!(S')$, font=\footnotesize] (theta) {$\bm{\theta}$\\Parameters};
			\node[box, below=1.2cm of S] (QFI_NH) {Precision remains elusive};
			\node[box, below=1.2cm of S'] (QFI) {$\mathcal{F}_{\theta}^\mrm{metric}$};
			
			\draw[->] (S) -- node[midway, below, font=\footnotesize] {Mapping $\mathcal{E}$} (S');
			\draw[->] (theta) -- (S);
			\draw[->] (theta) -- (S');
			\draw[->] (S') -- (QFI); 
			\draw[->] (S) -- (QFI_NH); 
			
		\end{tikzpicture}
	} 
	\caption{\textbf{A set of parameters }$\bm{\theta}$ is associated with non-Hermitian system $\mathcal{S}_\mrm{NH}$, where direct estimation of $\bm{\theta}$ is challenging. To simplify the process, $\mathcal{S}_\mrm{NH}$ is mapped, using a proper \emph{metric}, to a Hermitian system $\mathcal{S}_\mrm{H}$, which retains the \emph{same parameters} but allows for easier inference. The estimation of $\bm{\theta}$ in metric space is then achieved using the conventional QFI formalism for Hermitian systems.}
	\label{fig:ShSnh}
\end{figure}
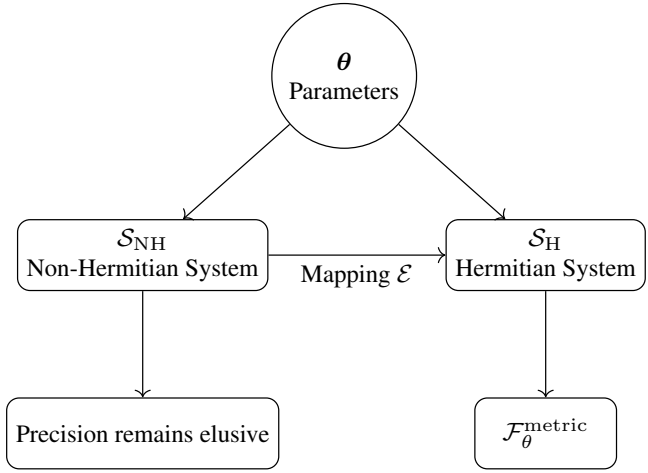

\subsection{General $2\times2$ non-Hermitian Hamiltonian}
Consider a non-Hermitian system $\mathcal{S}_\mrm{NH}(r,s,\tau,\phi)$ characterised by real parameters $r,s,\tau$, and $\phi$, through the following NHH~\cite{Mostafazadeh2003}
\begin{equation}\label{eq:general2x2}
	\tilde{H} = \begin{pmatrix}
							r + \tau C_{\phi} - \i\, s S_{\phi} & \i\, s C_{\phi} + \tau S_{\phi} \\
							\i \, s C_{\phi} + \tau S_{\phi} & r - \tau C_{\phi} + \i \,s S_{\phi}
						\end{pmatrix},
\end{equation}
where $C_{\phi} = \cos(\phi)$ and $S_{\phi} = \sin(\phi)$.  The eigenvalues of $\tilde{H}$ are real for $|s| \le |\tau|$. In \secref{sec:PhysSystems} we illustrate two different ways of calculating the metric operator $\eta$ in two physical systems, however, for $\tilde{H}$ in \eqnref{eq:general2x2}, we refer the reader to~\cite{Mostafazadeh2003}, where  the metric operator  and its square root are derived and are given  by 
\begin{align}
	\eta &= \begin{pmatrix}
					\sec(\alpha)    & \i \tan(\alpha) \\
					-\i\, \tan(\alpha) & \sec(\alpha)
				\end{pmatrix}, \nonumber\\ \mathcal{E} &= \sqrt{\eta} =  \begin{pmatrix}
																												r_{+} 		& -\i\, \, r_{-} \\
																												\i\, \, r_{-}  & r_{+}
																												\end{pmatrix},
\end{align}
with $	r_{\pm} := \frac{1}{2} \left( \sqrt{\sec(\alpha) - \tan(\alpha)} \pm \sqrt{\sec(\alpha) + \tan(\alpha)} \right)$.  The same system can be described by an equivalent Hermitian $\mathcal{S}_\mrm{herm}(r,s,\tau,\phi)$ with Hermitian  Hamiltonian
\begin{equation}
	H = \begin{pmatrix}
				r + \sqrt{ \tau^2 + s^2 } C_{\phi} & \sqrt{\tau^2 + s^2 } S_{\phi} \\
				\sqrt{\tau^2 + s^2 } S_{\phi} 		& r - \sqrt{\tau^2 + s^2 } C_{\phi}
			\end{pmatrix}.
\end{equation}
Since the new system depends on the same set of parameters $r, s, \tau$ and  $\phi$, we choose to estimate them in the Hermitian frame instead. For this, let us consider the following  initial state with real parameters $\theta$ and $x$:
\begin{equation}
	\rho_\mrm{in}= \bpm  1 - \theta   & x \\  x & \theta \epm,  \label{eq:initialstate}
\end{equation}
where $x \in [0, \sqrt{\theta(1-\theta)}]$.  Under the time dynamics, $\rho(t) = \exp(-\i H t ) \rho_\mrm{in} \exp(\i H t )$, estimating the parameters of the Hamiltonian requires calculating the QFI  matrix. This can be accomplished by applying the formula in \eqnref{eq:QFImixed} for mixed qubit state, resulting in the QFI
\begin{equation}
	\mathcal{F} = \bpm  
								0 &                0 					& 							0	& 			0 \\
							    0 & 	\mathcal{F}_{ss} & 	\mathcal{F}_{st}  & 	\mathcal{F}_{s\phi} \\
							    0 & 	\mathcal{F}_{ts} & 	\mathcal{F}_{tt} & 	\mathcal{F}_{t\phi} \\
								0  & 	\mathcal{F}_{\phi s} & 	\mathcal{F}_{\phi s} & 	\mathcal{F}_{\phi \phi}
	                         \epm,
\end{equation}
where 
	\begin{align}
	\mathcal{F}_{ss}  &=  \frac{4 s^2 t^2 [(2 \theta -1) S_{\phi} +2 x C_{\phi}]^2}{s^2+\tau ^2}, \nonumber \\
	  \mathcal{F}_{s\phi} &= \frac{s t S_{\alpha} \left[\kappa_{-}  S_{2\phi} + \lambda C_{2\phi} \right]}{\sqrt{s^2+\tau ^2}}, \nonumber \\
	\mathcal{F}_{s\tau} &= \frac{\tau}{s} \mathcal{F}_{ss}, \mathcal{F}_{\tau s}  =  \mathcal{F}_{s \tau},  \mathcal{F}_{\tau \tau}  =  \frac{\tau}{s}  \mathcal{F}_{\tau s}, \nonumber \\
	  \mathcal{F}_{\tau \phi}  &= \frac{\tau}{s}\mathcal{F}_{s \phi}, 	\mathcal{F}_{\phi s} = \mathcal{F}_{s \phi},  \mathcal{F}_{\phi \tau} = \mathcal{F}_{\tau \phi}, \nonumber \\
	\mathcal{F}_{\phi \phi} &= \frac{1}{2} S^2_{\alpha} \left[\kappa_{-}  C_{2\phi} - \lambda S_{2\phi} \right]  - \kappa_{+} S^2_{\alpha/2} (C_{\alpha}-3),
\end{align}
where various parameters are $\alpha \!=\! 2 t \sqrt{s^2 + \tau^2}$,  $\kappa_{\pm}  \!=\!  \left((1-2 \theta )^2 \pm 4 x^2\right)$ and  $\lambda  \!=\!  4 (2 \theta -1) x$.

In some cases the parameters of the input state iteself may be unknown, for example, the population $\theta$ of different energy levels or the coherence $x$ betweeen them. In that case, the relevant QFI matrix elements are given by 
\begin{align}
		\mathcal{F}_{\theta \theta} &= \frac{1 - 4 x^2}{\theta(1 - \theta)  - x^2}, \nonumber \\
		\mathcal{F}_{xx} &= \frac{4 (1 - \theta) \theta }{\theta(1 - \theta)  - x^2 }, \nonumber \\
		\mathcal{F}_{\theta x} &= \frac{2x (2 \theta - 1)}{\theta(1 - \theta)  - x^2} = \mathcal{F}_{x \theta}.	 \label{eq:QFIstate}
\end{align}
The QFI matrix elements presented here, encompassing both diagonal and off-diagonal components, are essential for characterizing  unknown parameters whether for both Hamiltonian and the input state. The diagonal elements capture the system's sensitivity to variations in individual parameters, thereby providing critical insights into the uncertainties linked to each parameter. In contrast, the off-diagonal elements elucidate the interdependencies between these parameters, demonstrating how fluctuations in one can significantly impact the estimation of another.

\subsection{Impact of quantum jumps on quantum Fisher information}

In light of \secref{sec:MasterEqn}, a crucial aspect of understanding QFI in non-Hermitian metrology is assessing the impact of quantum jumps on parameter sensitivity. One way to analyze this effect is by comparing the QFI obtained from the full master equation solution with that of the no-jump evolution. To facilitate this comparison, we examine the finite-difference form of \eqnref{eq:ME} for small $dt$, which systematically captures the influence of quantum jumps on the dynamics. Specifically, we express the evolution of the density matrix as  
\begin{align}
	\rho_\mrmb{me}(t + dt) &= \rho_\mrmb{me}(t) - \i\, [H, \rho_\mrmb{me}(t)] dt \nonumber \\ 
	&\quad +  \sum_{\ell} \left( \Gamma_{\ell} \rho_\mrmb{me}(t)  \Gamma_{\ell}^\dagger - \frac{1}{2}  \{ \rho_\mrmb{me}(t), \Gamma_{\ell}^\dagger \Gamma_{\ell} \} \right) dt \nonumber \\ 
	&\quad + \mathcal{O}(dt^2) \nonumber \\ 
	&= \rho_\mrmb{me}(t) - \i\,  \left( H_{\text{eff}} \rho_\mrmb{me}(t) - \rho_\mrmb{me}(t) H_{\text{eff}}^\dagger \right) dt \nonumber \\ 
	&\quad + dt \sum_{\ell} \Gamma_{\ell} \rho_\mrmb{me}(t) \Gamma_{\ell}^\dagger + \mathcal{O}(dt^2), \label{eq:MEdiff}
\end{align}
where $H_\mrmb{eff}$, the effective Hamiltonian, is given by \eqnref{eq:Heff}.  

To reformulate this evolution in a more insightful form, we define the non-unitary evolution operator  
\begin{equation}
	M_{0}(dt) = \mathbb{1} - \i\, H_\mrmb{eff} dt \approx e^{-\i H_\mrmb{eff} dt},
\end{equation}
which describes the no-jump evolution. Additionally, we introduce  
\begin{equation}
	M_{\ell}(dt) =  \Gamma_{\ell} \sqrt{dt},
\end{equation}
which accounts for the effect of quantum jumps. With these definitions, \eqnref{eq:MEdiff} can be rewritten as  
\begin{equation}
	\rho_\mrmb{me}(t + dt)  = M_{0} \rho_\mrmb{me}(t) M_{0}^\dagger  + \sum_{\ell > 0} M_{\ell} \rho_\mrmb{me}(t) M_{\ell}^\dagger +  \mathcal{O}(dt^2).
\end{equation}
These operators satisfy the completeness relation  
\begin{equation}
	\sum_{\ell} M_{\ell}^\dagger M_{\ell} = \mathbb{1} + \mathcal{O}(dt^2),
\end{equation}
which ensures proper normalization up to second-order corrections in $dt$.  

This decomposition of the master equation clearly distinguishes between contributions from the no-jump evolution and quantum jumps, providing a systematic way to analyze their respective effects on QFI. Specifically, the full density matrix can be expressed as a probabilistic mixture of conditional states:  
\begin{equation}
	\rho_\mrmb{me}(t + dt) = p_{0}(t + dt) \rho_\mrmb{nj}(t+dt) + \sum_{\ell} p_{\ell} (t + dt) \rho^{(\ell)}_\mrmb{j}(t + dt). \label{eq:MEcond}
\end{equation}
Here, the individual components are given by:  
\begin{itemize}
	\item \textbf{No-jump evolution:}  
	The state $\rho_\mrmb{nj}(t+dt)$ corresponds to the evolution without quantum jumps, governed by the effective NHH:  
	\begin{equation}
		\rho_\mrmb{nj}(t+dt) = \frac{M_{0}(dt) \rho_\mrmb{me}(t) M_{0}^{\dagger}(dt)}{p_{0}(t + dt)}, \label{eq:rho_nj}
	\end{equation}
	occurring with probability  
	\begin{equation}
		p_{0}(t + dt) = \Tr[M_{0}(dt)  \rho_\mrmb{me}(t) M_{0}^{\dagger}(dt) ]. \label{eq:p0}
	\end{equation}
	
	\item \textbf{Jump evolution:}  
	The states $\rho^{(\ell)}_\mrmb{j}(t + dt)$ represent the system's evolution conditioned on the occurrence of the $\ell$-th quantum jump
	
	\begin{equation}
		\rho^{(\ell)}_\mrmb{j}(t + dt) = \frac{M_{\ell} \rho_\mrmb{me}(t) M_{\ell}^\dagger}{p_{\ell}(t + dt)}, \label{eq:rho_ell}
	\end{equation}
	which occurs with probability  
	\begin{equation}
		p_{\ell}(t + dt) = \Tr[M_{\ell}(dt)  \rho_\mrmb{me}(t) M_{\ell}^\dagger(dt) ], \quad \ell > 0. \label{eq:pell}
	\end{equation}
\end{itemize}
This formulation explicitly separates the deterministic no-jump evolution from the stochastic jump-induced transitions, offering a clear framework to quantify their respective contributions to QFI. This decomposition plays a crucial role in establishing bounds on QFI, as~\cite{Ma2019}
\begin{equation}
		\mathcal{F}_{\theta}\left(\rho_\mrmb{nj}(t{+}dt)\right)  \ge \mathcal{F}_{\theta}\left(\rho^{(\ell)}_\mrmb{me}(t{+}dt)\right).
\end{equation}

\section{Physical systems}\label{sec:PhysSystems}
     We now examine two physical scenarios to illustrate the behavior of  QFI and the associated system properties. The first scenario involves a static metric operator, while the second considers a dynamic metric operator. These contrasting cases will help to highlight the differences in QFI and its implications for the system.
     
\subsection{Time-independent metric: Biorthogonal system}
 Consider the  non-Hermitian Hamiltonian 
 \begin{equation}
 	\tilde{H}_{1} = \begin{pmatrix} \omega_0 - \i\, \gamma & g \\ g & \omega_0 + \i\, \gamma \end{pmatrix},  \label{eq:H1nh}
 \end{equation}
	which  represents a two-level quantum system, where the levels are coupled through a parameter \(g\), and experience gain and loss dynamics characterized by the imaginary terms \(\pm \i \gamma\). The real part of the diagonal elements, \(\omega_0\), represents the energy levels of the two states, while the off-diagonal coupling term \(g\) dictates the strength of interaction between the states. The imaginary parts, \(-\i \gamma\) and \(+\i \gamma\), introduce non-Hermitian behavior, corresponding to the loss (for the first state) and gain (for the second state) mechanisms. This type of Hamiltonian is commonly encountered in systems exhibiting (\(\mathcal{PT}\))--symmetry, where the balance between loss and gain plays a critical role in determining the system's dynamics and phase transitions. Such models are often used in the study of open quantum systems, photonics, and systems where energy exchange with the environment is present.

	 The  eigenvalues of $\tilde{H}_{1}$ are $ \tilde{\lambda}_{\pm} = \omega_0 \pm \sqrt{g^2 - \gamma^2 }$ and are real for $g > \gamma$. The eigenvectors $\{\ket{\tilde{\psi}_n}, \ket{\tilde{\phi}_n} \}: \tilde{H}_{1} \ket{\tilde{\psi}_n} =  \tilde{\lambda}_{n} \ket{\tilde{\psi}_n}$, $\tilde{H}_{1}^{\dagger} \ket{\tilde{\phi}_n} =  \tilde{\mu}_{n} \ket{\tilde{\phi}_n}$ form a \emph{biorthogonal} system \cite{Brody2013}. In such biorthogoal systems, the metric can be constructed as  \cite{Mostafazadeh2002}
 \begin{equation}
 	\eta = | \tilde{\phi}_{1} \rangle \langle \tilde{\phi}_{1} |  + | \tilde{\phi}_{2} \rangle \langle \tilde{\phi}_{2} | =  \mathbb{1}  - \frac{ \gamma }{g} \sigma_{y}. \label{eq:etaDef}
 \end{equation}

  As expected, $\eta \to \mathbb{1}$ as $\gamma \to 0$, i.e., when $\tilde{H}_1$ approaches the Hermitian limit.
  By calculating the  positive square root of this metric operator 
\begin{equation}
		\E = \sqrt{\eta} = \frac{1}{2}\left( \Gamma_{+} +  \Gamma_{-} \right) \mathbb{1} -  \frac{1}{2}\left(\Gamma_{+} -  \Gamma_{-}\right) \sigma_{y},
\end{equation}
where $\Gamma_{\pm} = \sqrt{1 \pm  \frac{\gamma}{g}}$,  we obtain the Hermitian counterpart of  $\tilde{H}_1$  which is given by 
\begin{equation}
	H_{1} = \E \tilde{H}_{1} \E^{-1} = \bpm   \omega_{0}  &  \sqrt{g^2 - \gamma^2}  \\  \sqrt{g^2 - \gamma^2}  & \omega_{0} \epm.   \label{eq:HermH1}
\end{equation}
Note that the  eigenvalues  of  $H_{1}$ are $\lambda_{\pm}=\omega_{0} \pm \sqrt{g^2 - \gamma^2}$ same as that of $\tilde{H}_{1}$ but the eigenvectors $\ket{\psi_{\pm}} = \bpm \pm 1 & 1 \epm^\T$ are \emph{orthogonal}.

We now turn to the dynamics governed by $\tilde{H}_1$, where we examine the three approaches outlined in \secref{sec:approaches}, namely the  \emph{metric}, \emph{naive--normalization} and \emph{master equation} approaches.

We begin with the metric formalism, where the dynamics is governed by the Hermitian counterpart of the Hamiltonian, $\tilde{H}_{1}$, as obtained in \eqnref{eq:HermH1}. This Hamiltonian evolves the qubit state in \eqnref{eq:initialstate} as follows:
\begin{equation}
	\rho_\mrm{metric}(t) = e^{-\i H t} \rho(0) e^{\i H t} =  \sum\limits_{\ell =0}^{3} c_{\ell} \sigma_{\ell},    \label{eq:rho1metric}
\end{equation}
with $c_{0} = 1/2$, $c_{1} = x$, $c_{2} = (\theta - \frac{1}{2}) \sin (2t\sqrt{g^2 - \gamma^2})$ and $c_{3} = (\theta - \frac{1}{2}) \cos (2t\sqrt{g^2 - \gamma^2})$.

On the other hand, the naive--normalization approach involves writing the Schr\"{o}dinger equation for $\tilde{H}_{1}$ with the solution
\begin{equation}
\tilde{\rho}(t) = e^{-\i\tilde{H}_{1}t} \tilde{\rho}(0) e^{\i \tilde{H}_{1}^\dagger t}, \label{eq:tildrhot}
\end{equation}
which is a trace non-preserving equation. For example, with the  initial state given in \eqnref{eq:initialstate}, the trace of $\rho_\mrm{in}$ and $\tilde{\rho}(t)$ with respect to the \emph{standard} metric is:
\begin{align}
		\Tr[\rho_\mrm{in}] =& 1, \nonumber \\
{\rm and}\quad 		\Tr[\tilde{\rho}(t)] = &\frac{\gamma  (2 \theta -1)}{\sqrt{g^2-\gamma^2 }}  S_{\beta}  -\frac{\gamma ^2}{g^2-\gamma^2 } C_{\beta} +\frac{g^2}{g^2-\gamma^2 },
\end{align}
where $\beta = 2 t \sqrt{g^2-\gamma ^2}$. Clearly, the trace is not preserved for any time $t > 0$. To make \eqnref{eq:tildrhot} physically meaningful, one may divide it by its trace, yielding the following \emph{normalized} density matrix:
\begin{equation}
	\tilde{\rho}_\mrm{norm}(t)  = \frac{\tilde{\rho}(t) }{\Tr[\tilde{\rho}(t) ]}. \label{eq:rhoNorm}
\end{equation}
Note that, in the \emph{metric} space with the metric operator given in \eqnref{eq:etaDef}, we have from \eqnref{eq:avg}, $ \Tr[\eta \tilde{\rho}(t) ] = 1$ -- demonstrating the preservation of trace of $\tilde{\rho}(t)$ in \eqnref{eq:tildrhot} in the metric space for all the evolution times and, hence, the unitary nature of dynamics.

\begin{figure}
	\makebox[80mm]{\large \textbf{(a)}}  
	\includegraphics[width=\linewidth]{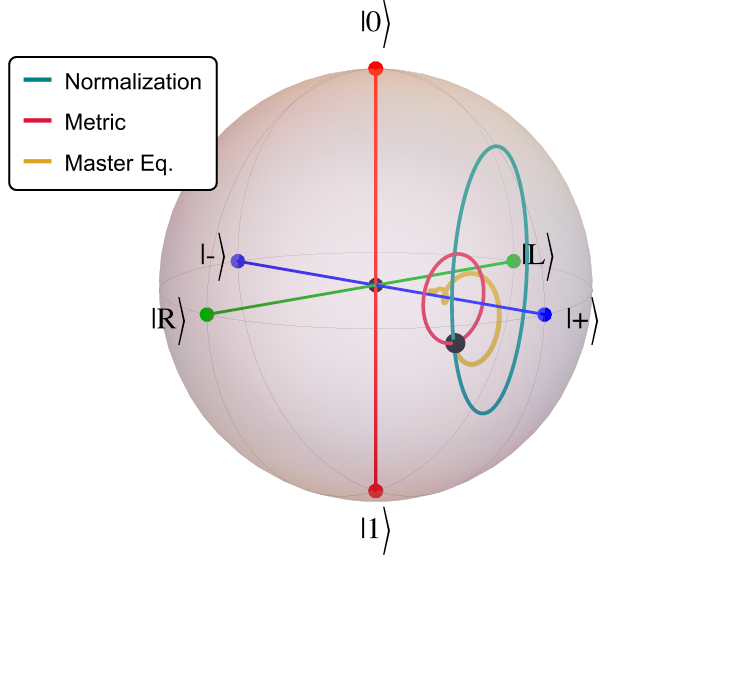} 
	\makebox[80mm]{\large \textbf{(b)}}  
	\includegraphics[width=\linewidth]{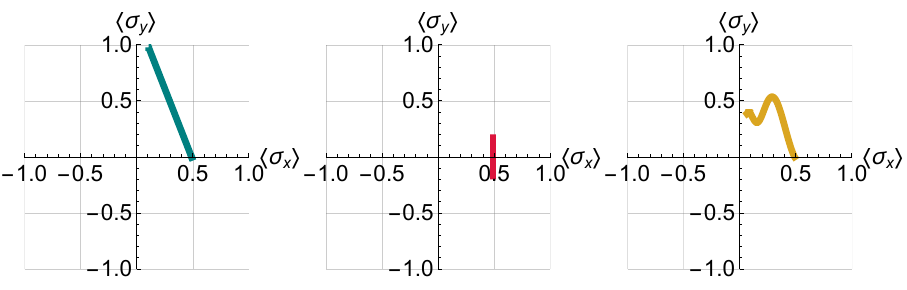}
	\caption{ \textbf{(a)} \textbf{Evolution of different states in the Bloch sphere}:  $\rho_\mrm{metric}(t)$ in \eqnref{eq:rho1metric},  time normalized  $\tilde{\rho}_\mrm{norm}(t)$  in \eqnref{eq:rhoNorm}, and  $\rho_\mrm{me}(t)$ -- the solution of the master equation \eqnref{eq:rhoME}. The initial state corresponds to Bloch vector $(0.48,0,-0.2)$ and pertains to  $\theta = 0.6$ and $x = 0.24$ in \eqnref{eq:initialstate}. The Hamiltonian parameters used are $g = 0.5$ and $\gamma = 0.4$. \textbf{(b)} \textbf{The  evolution projected }onto the Bloch sphere equator.
	              }
	\label{fig:BlochSphere}
\end{figure}

Finally, the NHH  \eqnref{eq:H1nh} can be interpreted as an effective Hamiltonian emerging after neglecting the jump term in the following master equation 
\begin{align}
	\partial_{t} \rho_\mrm{me}(t) = &-\i\, [H, \rho_\mrm{me}(t)  ] +  2 \gamma (\sigma_{-}  \rho_\mrm{me}(t) \sigma_{+} \nonumber  \\&- \frac{1}{2} \sigma_{+} \sigma_{-}  \rho_\mrm{me}(t)  - \frac{1}{2} \rho_\mrm{me}(t) \sigma_{+} \sigma_{-}).  \label{eq:rhoME}
\end{align}
where $H = \omega_{0}  \mathbb{1} + g \sigma_{x}$. the term $\omega_{0}  \mathbb{1}$ adds a constant global energy shift, which does not affect the system's dynamics, while the term $g \sigma_{x}$ generates coherent oscillations (rotation around the $x$--axis).  The system attains a steady state as $t \to \infty$ given by
\begin{equation}
	\rho_\mrm{me,ss} = \frac{g^2}{2(g^2 - \gamma^2)} \bpm    1    &  -\i\, \gamma/g  \\  \i\, \gamma/g   &  1 + (\gamma/g)^2     \epm.
\end{equation}
The system's dynamics under different formulations---namely, the metric-based evolution $\rho_\mrm{metric}(t)$, the normalized state evolution $\tilde{\rho}_\mrm{norm}(t)$, and the master equation solution $\rho_\mrm{me}(t)$---are illustrated in~\figref{fig:BlochSphere}. These trajectories, shown for a representative initial state on the Bloch sphere, highlight the distinctions and similarities among the three approaches.  Figure~\ref{fig:BlochSphere} (b) further provides a projection of this evolution onto the equatorial plane, offering additional geometric insight.

\emph{Quantum Fisher Information}: To elucidate how the aforementioned approaches can yield different results in terms of the QFI, we will examine the input state defined in \eqnref{eq:initialstate} and concentrate on the estimation of the parameter \(\theta\). By computing the QFI, as outlined in  \eqnref{eq:QFImixed}, we can quantitatively assess the effectiveness of each approach. The normalization and metric approaches respectively lead to 
\begin{align}
\mathcal{F}^\mrm{norm}_{\theta} &= \frac{\left(g^2-\gamma ^2\right)^2 	\mathcal{F}^\mrm{metric}_{\theta}}{ g^2 + \gamma  (2 \theta -1) \sqrt{g^2 - \gamma^2} S_{\beta} - \gamma ^2  C_{\beta}}, \label{eq:QFInorm} \\
\mathcal{F}^\mrm{metric}_{\theta} &=  \frac{1 - 4 x^2}{\theta(1 - \theta)  - x^2}. \label{eq:QFImetric}
\end{align}
Thus,  on one  hand we have a precision (QFI) which is constant for a given value of state parameter  $x$ under metric formalism, the normalization method predicts an oscillating behavior with time.  It should be noted that $\mathcal{F}_{\theta}^\mrm{metric}$ in \eqnref{eq:QFImetric} is indeed the first element of QFI matrix in \eqnref{eq:QFIstate} derived  for the general $2 \times 2$ NHH.

\begin{figure}
\makebox[80mm]{\large \textbf{(a)}}  
\includegraphics[width=80mm]{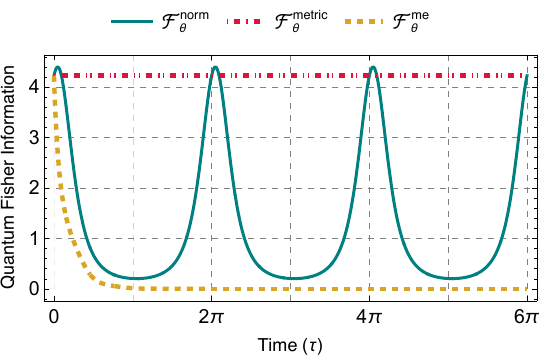}  
 \makebox[80mm]{\large \textbf{(b)}} 
\includegraphics[width=80mm]{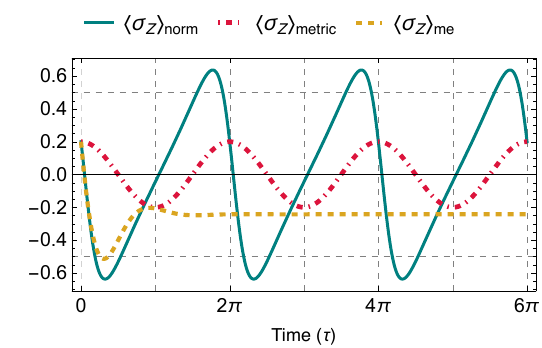}
	\caption{ \textbf{(a)} \textbf{QFI as a function of} $\uptau = t/2\sqrt{g^2 - \gamma^2}$, corresponding to parameter $\theta$ as given  $\mathcal{F}^\mrm{norm}_{\theta}$ in \eqnref{eq:QFInorm}, $\mathcal{F}^\mrm{metric}_{\theta}$  in  \eqnref{eq:QFImetric}, $\mathcal{F}^\mrm{me}_{\theta}$ for the state obtained as solution of \eqnref{eq:rhoME} -- choosing parameters $\theta = 0.6$, $x=0.24$, $g= 0.5$, $\gamma = 0.4$. \textbf{(b)} \textbf{Expectation value} of $\sigma_{z}$ operator for different states. }
	\label{fig:QFI}
\end{figure}

For comparison, we obtain the QFI, $\mathcal{F}^\mrm{me}_{\theta}$ from the solution of the master equation in \eqnref{eq:rhoME} and  plot it in \figref{fig:QFI} along with $\mathcal{F}^\mrm{norm}_{\theta}$ and $\mathcal{F}^\mrm{metric}_{\theta}$  given in Eqs.~(\ref{eq:QFInorm}) and (\ref{eq:QFImetric}), respectrively.  In the limit, where the decay rate $\gamma \to 0$ [i.e., the Hermitian limit in \eqnref{eq:H1nh}], all QFI measures converge to their corresponding Hermitian counterparts
\begin{equation}
	\mathcal{F}^\mrm{norm}_{\theta}|_\mrm{\gamma = 0}   = \mathcal{F}^\mrm{metric}_{\theta}|_\mrm{\gamma = 0} = \mathcal{F}^\mrm{me}_{\theta}|_\mrm{\gamma = 0} = \frac{1 - 4 x^2}{\theta(1 - \theta)  - x^2}.  \label{eq:gammaZero}
\end{equation}
An interesting comparison arises from \eqnref{eq:QFInorm} when considering the QFI in the normalization formalism, $\mathcal{F}^\mathrm{norm}_{\theta}$, and its metric counterpart, $\mathcal{F}^\mathrm{metric}_{\theta}$:
\begin{equation}
	\mathcal{F}^\mathrm{norm}_{\theta} = \frac{1}{\Tr[\tilde{\rho}(t)]^2} \mathcal{F}^\mathrm{metric}_{\theta}.
\end{equation}
This expression shows that the QFI in the normalization formalism is scaled by the factor $\frac{1}{\Tr[\tilde{\rho}(t)]^2}$ relative to the metric frame. Notably, when $\Tr[\tilde{\rho}(t)] < 1$, the normalization formalism can lead to an enhancement in QFI, surpassing the Hermitian limit. This seems to imply  that the normalization formalism offers a potential advantage in quantum parameter estimation when the trace of the non-Hermitian density matrix is less than unity, reflecting a \emph{spurious} broader range of information captured by this approach.

 Further, from \figref{fig:QFI} we have  contrasting predictions with time about precision in measuring parameter $\theta$.  Under the metric formalism, one can always be sure of obtaining information about parameter $\theta$ and this information does not change over time. However, the normalization method suggests that the precision would depend on the time of measurement. In particular, one can beat the precision from metric formalism by performing measurements at  $t = 4n \pi/\sqrt{g^2 - \gamma^2}$, for a give value of $g$ and $\gamma$ with $g > \gamma$ [as already assumed above in order to ensure that the eigenvalues are real].

\subsection{Time--dependent metric }\label{subsec:tdm}
\begin{figure}
	\makebox[80mm]{\large \textbf{(a)}}  
	\includegraphics[width=\linewidth]{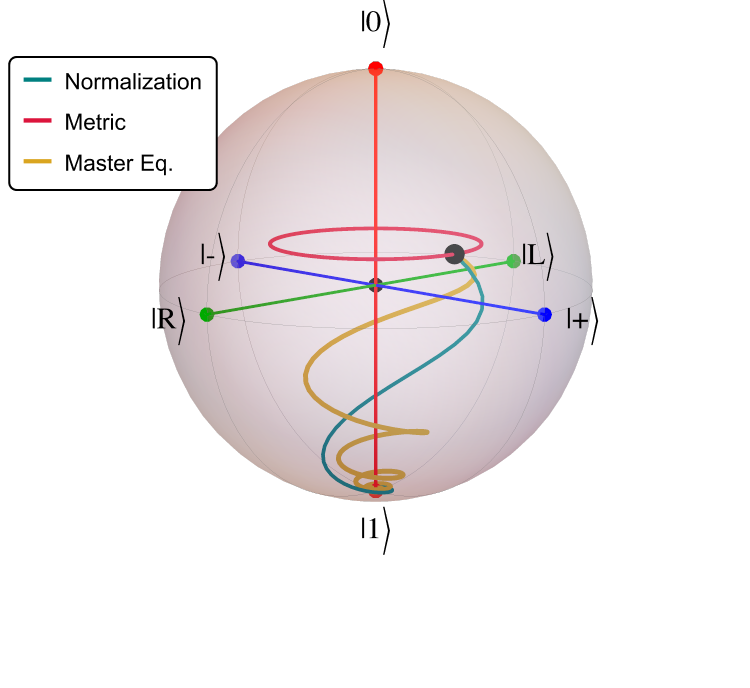} 
	\makebox[80mm]{\large \textbf{(b)}}  
	\includegraphics[width=\linewidth]{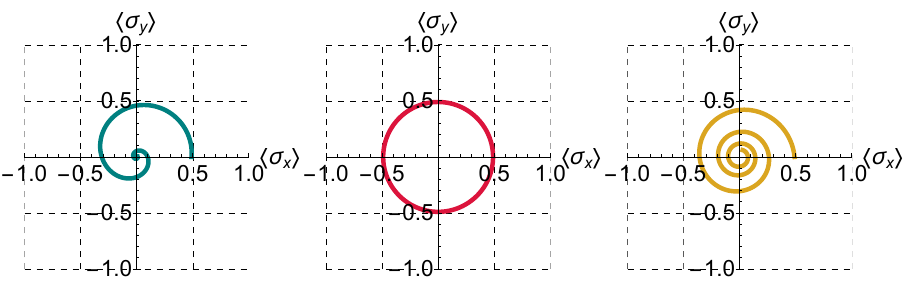}
	\caption{ \textbf{(a)} \textbf{Evolution of different states in the Bloch sphere}:  The states shown are the same as in \figref{fig:BlochSphere} but this time in the context of the NHH  given in \eqnref{eq:Hnh2}. The initial state  pertains to  $\theta = 0.4, x = 0.24$. \textbf{(b)} \textbf{The  evolution projected} onto the Bloch sphere equator.}
	\label{fig:BlochSphere2}
\end{figure}

Next, we consider the Hamiltonian $H = \omega \sigma_z$, which describes the coherent evolution of the system, leading to precession about the $z$ axis of the Bloch sphere \figref{fig:BlochSphere2} (a), and the dissipation is taken into accoount by Lindblad operator $L = \sqrt{\gamma} \sigma_{-}$.  This model captures the spontaneous emission process, where the qubit relaxes from the excited state $\ket{0}$ to the ground state $\ket{1}$ with a rate $\gamma$. This describes energy loss, such as photon emission in a two-level atom. When  $\gamma \gg \omega$, dissipation dominates, leading to rapid relaxation and loss of coherence. For $\gamma \ll \omega$, coherent oscillations last longer before dissipation takes over. This balance is crucial in quantum optics and information processing for controlling decoherence.  The effective Hamiltonian is given by 
\begin{equation}
	\tilde{H}_\mrm{2} = H -  \frac{\i}{2} \Gamma^\dagger \Gamma = (\omega - \i\, \gamma) \sigma_{z}, \label{eq:Hnh2}
\end{equation}
 up to a constant proportional to identity. Unlike $\tilde{H}_{1}$ in \eqnref{eq:H1nh} of the previous example, the eigenvectors of $\tilde{H}_\mrm{2}$ are orthogonal, hence the method of constructing the metric for biorthogonal systems, as given in \eqnref{eq:etaDef}, does not apply.  One can however obtain the metric in this case by solving \eqnref{eq:metricEOM} which yields 
 \begin{equation}
 	\eta(t) = \exp(2\gamma t \sigma_z)=\mathbb{1} \cosh(2\gamma t) + \sigma_{z} \sinh(2\gamma t).
 \end{equation}
 This leads to the Hermitian equivalent of  $\tilde{H}_\mrm{2}$ given by $H_2 = \omega_{0} \sigma_z$.   Without discussing the dynamics at length like in the previous section, one can similarly obtain the $\rho_\mrm{metric}$, $\tilde{\rho}_\mrm{norm}$, and $\rho_\mrm{me}$. 

We now consider  a general observable $A = \sum\limits_{\ell =1}^{3}  a_\mrm{\ell } \sigma_\mrm{\ell}$,  with $a_\mrm{\ell } = \frac{1}{2} \Tr[A \sigma_\mrm{\ell}] $, $\mrm{\ell} = 1,2,3$. Using again the initial state \eqnref{eq:initialstate}, we obtain
	\begin{align}
	\langle A \rangle_\mrm{metric} = &(1 -2 \theta)  a_{3} + 2 x [1 -  \cos(2\omega_{0} t)] a_{1} \nonumber \\&+ 2 x  \sin(2\omega_{0} t) a_\mrm{2}, \\
	\langle A \rangle_\mrm{norm}   &= \alpha_{0}  \langle A \rangle_\mrm{metric} + \alpha_1 a_3, \\
	\langle A \rangle_\mrm{me} & = \beta_{0} \langle A \rangle_\mrm{norm}  +  \beta_{1}  a_{3},
\end{align}
where $\langle A \rangle_{\bullet} = \Tr[A \rho_{\bullet}]$ with $\bullet = \mrm{norm}, \mrm{metric}, \mrm{me}$ and
\begin{align}
	\alpha_{0} &= \frac{e^{2 \gamma  t}}{\theta  \left(e^{4 \gamma  t}-1\right)+1} = \frac{1}{\Tr[\tilde{\rho}(t)]},\\
	\alpha_1 &= \frac{\left(e^{2 \gamma  t}-1\right) \left[\theta  \left(e^{2 \gamma  t}-1\right)+1\right]}{\theta  \left(e^{4 \gamma  t}-1\right)+1},\\
	\beta_{0} &=  \theta + (1- \theta)e^{-4 \gamma t},\\
	\beta_{1} &= -(1-\theta)(1-e^{-4\gamma t}).
\end{align}
In the long time limit, both the normalization as well as the master equation approaches show that the system settles at the ground state $\ket{1}$, as shown in  \figref{fig:BlochSphere2}, and the average of the observable $A$ approaches the following limit
\begin{align}
	\lim_{t \to \infty} \langle A \rangle_\mrm{norm}  = -a_{3} = \lim_{t \to \infty} \langle A \rangle_\mrm{me}.
\end{align}
However, as expected, in the metric frame there is no such decohering effects as the system evolves under unitary dynamics.
The QFI corresponding the parameter $\theta$ follow similarly as discussed in the previous example. We obtain
\begin{align}
			\mathcal{F}_\mrm{\theta}^\mrm{metric} &= \frac{1 - 4 x^2}{\theta (1 - \theta)  - x^2} \label{eq:QFImetric2},\\
					\mathcal{F}_\mrm{\theta}^\mrm{norm} &=  \frac{\left(1 - 4 x^2\right) e^{4 \gamma  t}}{\left[\theta (1 - \theta) - x^2\right] \left[\theta  \left(e^{4 \gamma  t}-1\right)+1\right]^2} \nonumber  \\&=   \frac{e^{4 \gamma  t}}{\left[\theta  \left(e^{4 \gamma  t}-1\right)+1\right]^2}  \mathcal{F}_\mrm{\theta}^\mrm{metric} = \frac{1}{\Tr[\tilde{\rho}(t)]^2} \mathcal{F}_\mrm{\theta}^\mrm{metric} \label{eq:QFInorm2}, \\
			\mathcal{F}_\mrm{\theta}^\mrm{me} &= \frac{e^{-2 \gamma  t} \left(4 x^2-e^{2 \gamma  t}\right)}{(1 - \theta)^2+e^{2 \gamma  t} \left(\theta +x^2-1\right)}.  \label{eq:QFIme2}
\end{align}
The metric approach once again demonstrates its effectiveness by predicting a constant, time-independent precision in quantum measurements of parameter $\theta$. This stability suggests that the choice of metric plays a crucial role in maintaining consistent precision over time, thereby providing a reliable framework for parameter estimation.

\begin{figure}
	\makebox[80mm]{\large \textbf{(a)}}  
	\includegraphics[width=80mm]{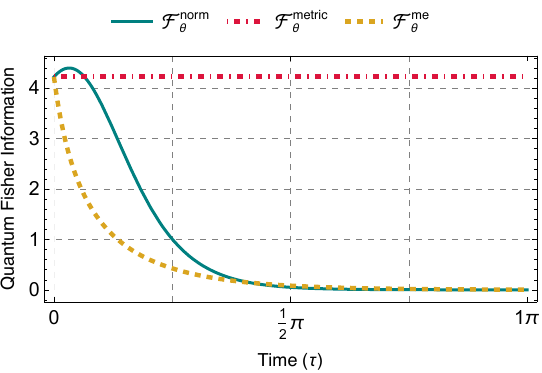}
	\makebox[80mm]{\large \textbf{(b)}}  
	\includegraphics[width=80mm]{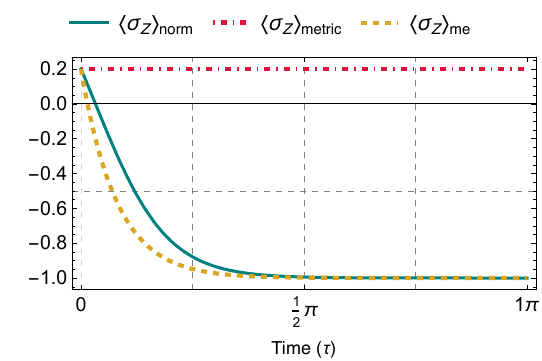}
	\caption{ \textbf{(a)}  \textbf{QFI as a function of} $\uptau =  t/\gamma$ corresponding to parameter $\theta$ as given  by $\mathcal{F}^\mrm{metric}_{\theta}$  in  \eqnref{eq:QFImetric2}, $\mathcal{F}^\mrm{norm}_{\theta}$ in \eqnref{eq:QFInorm2}, $\mathcal{F}^\mrm{me}_{\theta}$  in \eqnref{eq:QFIme2} -- the state parameters chosen are $\theta = 0.4$ and $x=0$, while the Hamiltonian parameters $g= 0.5$, $\gamma = 0.4$ are used.  \textbf{(b)} \textbf{Expectation value} of $\sigma_{z}$ operator for different states. }
	\label{fig:QFI2}
\end{figure}

In contrast, the normalization method, unlike in previous example, does not exhibit any recurring behavior. But it does indicate an enhancement in precision compared to the metric (Hermitized) case for, at least, a finite initial duration. This spurious improvement implies that the normalization method may produce results that warrant further scrutiny, as its effectiveness in enhancing quantum precision may not be universally applicable.

Additionally, the master equation approach reveals a more gradual decline in the QFI over time, ultimately trending towards zero. This behavior signifies a loss of information suggesting that, as the system evolves, the ability to extract precise measurements diminishes. The gradual decrease in the QFI underscores the importance of considering different approaches when evaluating quantum precision, as each method provides distinct insights into the dynamics of the system and its associated measurement capabilities.  The behavior of the QFI for the parameter $\theta$ is shown in~\figref{fig:QFI2} (a), comparing $\mathcal{F}^\mrm{metric}_{\theta}$, $\mathcal{F}^\mrm{norm}_{\theta}$, and $\mathcal{F}^\mrm{me}_{\theta}$ as functions of the dimensionless time parameter $\uptau = t/\gamma$. These results illustrate how different formulations capture the sensitivity of parameter estimation under the same initial state and Hamiltonian conditions. Figure~\ref{fig:QFI2} (b) complements this by displaying the expectation value of the $\sigma_z$ operator for the corresponding state evolutions, providing further dynamical insight.

Overall, these findings illustrate the nuanced differences among various approaches in predicting quantum measurement precision, emphasizing the need for careful selection of methodologies based on the specific characteristics of the quantum systems under investigation.

\section{Conclusion}\label{sec:Conclusion}
In this work, we have carried out a comparative study of different probability-conserving approaches to quantum parameter estimation in systems governed by non-Hermitian Hamiltonians (NHHs). Specifically, we focused on the \emph{metric formalism} and the widely used \emph{normalization approach}, while also including the \emph{master-equation framework} as the main reference for open system results. All three methods restore probability conservation in different ways, yet their conceptual foundations and physical implications differ substantially.

The normalization or naive approach, where evolution is  effected by propagating in time via a given non-Hermitian Hamiltonian, consists of reinforcing probability conservation by renormalizing the state at each time step. Statistical properties of the evolved states, such as observable averages or outcome probabilities then are extracted by assuming a trivial metric which is identical to that of Hermitian space.  
Although this ad hoc method often reproduces results close to those obtained from the corresponding master equation in weakly excited regimes, it can lead to inconsistencies and misleading conclusions in general cases—particularly by neglecting quantum jumps and the proper geometric structure of the underlying Hilbert space. As such, this naive normalization can overestimate metrological precision and even suggest apparent violations of fundamental bounds.

We demonstrate that the metric formalism provides a consistent and physically meaningful way in estimating parameters of an input state that evolves under a NHH. In such cases, by introducing a non-trivial metric operator (i.e. a suitable redefinition of the scalar product), a pseudo-Hermitian system can be mapped to an equivalent Hermitian, that is isolated, quantum system. This ensures that quantities such as the quantum Fisher information retain their standard operational meaning, allowing parameter estimation to be performed using conventional metrological tools without ambiguity.

The master-equation formalism, while computationally more demanding, naturally incorporates both coherent evolution and dissipative effects, thereby serving as a reliable benchmark for assessing the physical validity of approximate approaches. 

Taken together, our findings highlight that the metric formalism offers the most principled framework for performing quantum metrology in non-Hermitian systems. It ensures probability conservation, preserves the structure of quantum mechanics, and allows for consistent comparison with standard Hermitian results. The normalization method, though convenient, should therefore be applied with caution and only within regimes where its assumptions remain valid. This comparative analysis clarifies  ambiguities in non-Hermitian quantum metrology and provides a unified foundation for future studies of realistic open-system dynamics and time-dependent metric structures.

After completion of this work, we became aware of Ref.~\cite{arkhipov2025upper}, which proposes a covariant-derivative–based definition for evaluating the QFI directly from pseudo-Hermitian Hamiltonians. In the present study, we computed the QFI using the Hermitized Hamiltonian. Since the two approaches are physically equivalent, the covariant QFI is necessarily bounded from above by the maximum QFI in the Hermitized frame, as also noted in Ref.~\cite{arkhipov2025upper}.
\section*{Acknowledgements}
This work was supported by the Polish National Science Centre (NCN) under the Maestro Grant No. DEC-2019/34/A/ST2/00081. J.K.~acknowledges also the the support of the National Science Centre, Poland under the SONATA BIS grant no. 2023/50/E/ST2/00457.

%

\appendix
\setcounter{secnumdepth}{3}
\section{Standard vielbein formalism in general relativity}\label{app:vielbein}

Here we provide a brief reminder of the meaning and function of a
vielbein within the standard vielbein formalism.

A vielbein (from the German ``many legs'') is a geometric object
widely used in differential geometry and general relativity
\cite{Nakahara2003, Carroll2019}. It is also known as a frame
field, and is sometimes described as the ``square root'' of the
metric. In specific dimensions, it is referred to as a tetrad or
vierbein in four dimensions, and a triad or dreibein in three
dimensions.

In general relativity, space-time is described by a metric
$g_{\mu\nu}(x)$, which defines distances and angles in a curved
space. However, in a curved space, it is often useful to work with
a locally flat coordinate system at each point. This is precisely
the role of the vielbein: it provides a local orthonormal frame at
each point of a curved manifold, thereby relating the curved
spacetime metric to a locally flat (Minkowski) metric. This
correspondence greatly simplifies calculations, especially when
dealing with spinor fields and local Lorentz symmetry.

In this sense, a vielbein plays a role closely analogous to that
of a Dyson map \cite{Dyson1956, Ghosh2025} in non-Hermitian
quantum mechanics, which is the focus of our paper: both serve as
structure-preserving transformations that relate a ``curved'' or
non-orthonormal description to a locally orthonormal one with a
simplified metric. The Dyson map is sometimes referred to as a
generalized vielbein \cite{Ju2022, Ju25}, and we adopt this
convention here.

More formally, let us consider the inner product between two
vectors, ${u}$ and ${v}$, in a curved space described by the
metric $g_{\mu\nu}(x)$:
\begin{equation}
	\left< {u}, {v} \right> = u^{\mu} g_{\mu\nu} v^{\nu},
	\label{A1}
\end{equation}
where $g_{\mu\nu}(x)=g_{\nu\mu}(x)$ and the Einstein summation
rule is applied. The vielbein is usually defined as a set of
objects $e_\mu^{a}(x)$ that relates the curved space-time metric
$g_{\mu\nu}(x)$ to the flat (Minkowski) metric $\eta_{ab}$ via:
\begin{equation}
	g_{\mu\nu}(x)=e_{\mu}^{a}(x)\,\eta_{ab} \,e_{\nu}^{b}(x),
	\label{A2}
\end{equation}
where $\mu, \nu = 0, 1, ..., (d-1)$ are space-time indices, $a, b
= 0, 1, ..., (d-1)$ are local Lorentz (flat) indices, and
$e_\mu^{a}$ is the vielbein. The flat Minkowski metric is
$\eta_{ab}={\rm diag}([-1,1,1,1])$ in four-dimensional spacetime;
yet, for the sake of simplicity, one can set it equal to the
Kronecker delta $\delta_{ab}$. By redefining the components of the
vectors $u$ and $v$ as follows:
\begin{equation}
	\widetilde{{u}}^a = e_\mu^{a} {u}^\mu \quad {\rm and}
	\quad \widetilde{{v}}^b = e_\nu^{b} {v}^\nu,
	\label{A3}
\end{equation}
one obtains
\begin{equation}
	\left< {u}, {v} \right> = \widetilde{{u}}^a \eta_{a b}
	\widetilde{{v}}^b,
	\label{A4}
\end{equation}
which formally takes the same form as the inner product in a flat
space.  These relations show that the vielbein relates
(``translates'') curved coordinates and locally flat coordinates.
Intuitively, one can think that $g_{\mu\nu}$ specifies distances
and intervals in curved space-time, while $e_\mu^{a}$ determines
how to project vectors in curved space onto a local flat frame.
Thus, in a local neighborhood, the dynamics reduce to those of
special relativity defined on the tangent space, and the vielbein
formalism enables an explicit correspondence between curved-space
and locally inertial descriptions. Note that there is also an
inverse vielbein $e_a^{\mu}$, satisfying:
\begin{equation}
	e_a^{\mu} e_{\mu}^{b} = \delta_a^b, \qquad e_{\mu}^{a} e_a^{\nu} =
	\delta_{\mu}^{\nu}.
	\label{A5}
\end{equation}

Vielbeins are commonly applied in general relativity, especially
in describing coupling of spinor fields (like electrons) to
gravity \cite{Carroll2019}. Spinors need flat space gamma
matrices, so vielbeins are necessary. Moreover, they are commonly
applied in supergravity \cite{Freedman2012} (for defining, e.g.,
the spin connection) and in gauge gravity theories
\cite{Blagojevic2002}, where vielbeins appear as gauge fields of
translations.

Employing the vielbein formalism leaves the physical and geometric
predictions unchanged, while offering a more transparent and
conceptually illuminating formulation of the theory. There are,
however, important cases where the metric formulation fails and
only the vielbein formalism can be used, e.g., in the
above-mentioned gravitational theories that include fermions,
which require a locally defined orthonormal frame.

 \end{document}